\documentclass[draft,prb,twocolumn,floatfix,twoside,tbtags,showpacs]{revtex4-1}

\usepackage[utf8]{inputenc}
\usepackage{subfigure}
\usepackage{graphicx}
\usepackage{amsmath,amssymb}
\usepackage{bm}	
\usepackage{xspace} 

\newcommand{\etal}{\textit{et al}.\@\xspace}

\newcommand{\cf}{\textit{cf}.\@\xspace}

\newcommand{\md}{\mathrm{d}}
\newcommand{\mi}{\mathrm{i}}

\begin{document}

\title{Dislocation core field. I. Modeling in anisotropic linear elasticity theory}

\author{Emmanuel \surname{Clouet}}
\email{emmanuel.clouet@cea.fr}
\affiliation{CEA, DEN, Service de Recherches de Métallurgie Physique,
91191 Gif-sur-Yvette, France}


\date{\today}
\begin{abstract}
	Aside from the Volterra field, dislocations create a core field, 
	which can be modeled in linear anisotropic elasticity theory 
	with force and dislocation dipoles.
	We derive an expression of the elastic energy of a dislocation
	taking full account of its core field and show that no cross 
	term exists between the Volterra and the core fields.
	We also obtain the contribution of the core field 
	to the dislocation interaction energy with an external stress,
	thus showing that dislocation can interact with a pressure.
	The additional force that derives from this core field contribution
	is proportional to the gradient of the applied stress.
	Such a supplementary force on dislocations may be important
	in high stress gradient regions, such as close to a crack tip
	or in a dislocation pile-up.
\end{abstract}
\maketitle

\section{Introduction}

Far from a dislocation, the elastic field is well described 
by the Volterra solution \cite{Hirth1982}: this predicts that 
the displacement and the stress are respectively varying 
proportionally to the logarithm and the inverse of the distance
to the dislocation.
But the elastic field may deviate
from this ideal solution close to the dislocation.
Such a deviation corresponds 
to the dislocation core field.
It arises from anharmonicities in the crystal elastic behavior,
especially in the high-strained region of the core, 
as well as from perturbations
caused by the atomic nature of the core.
It is, in part, responsible for the dislocation formation 
volume, which manifests itself experimentally\cite{Crussard1949} 
through an increase of the lattice parameter with the dislocation density.
This leads to an interaction of the dislocation with an external pressure \cite{Hirth2005}.
Although the core field decays more rapidly than the Volterra field, 
it can modify the elastic interaction of dislocations with other defects
\cite{Henager2004,Henager2005,Clouet2009b}.
For instance, equilibrium distances in a dislocation pile-up
are affected by this core field\cite{Kuan1976}. As a consequence, the stress concentration 
at the tip of the pile-up is enhanced. This can favor fracture initiation
or yielding for an edge or mixed dislocation pile-up, 
or cross-slip for a screw pile-up\cite{Kuan1976}.
The stress produced by this core field also tends to open the crack in front of 
the dislocation pile-up on the glide plane\cite{Hirth1993b}. This explains the nucleation
of a crack in a mixed mode I-II or I-III sometimes observed experimentally in the glide plane 
of the pile-up. Without the core field, only modes II and III would be possible.
The core field can also alter dislocation properties such as their elastic energy\cite{Gehlen1972,Clouet2009b} 
or their dissociation distance in fcc metals \cite{Henager2004,Henager2005}.
Finally, it contributes to the elastic interaction between dislocations and impurities,
and thus may explain part of the solid solution hardening \cite{Fleischer1963}.

One can use atomistic simulations, either based on empirical potentials
or on \emph{ab initio} calculations, 
so as to take full account of the core field
when studying dislocations. 
On the other hand, the core field can be modeled 
within linear elasticity theory, 
using an equilibrated distribution of line forces
parallel to the dislocation and located close to its core
\cite{Gehlen1972,Hirth1973,Hoagland1976,Sinclair1978}. 
A multipole expansion of the distribution leads then to an
expression of the core field in term of a series. 
Usually, only the leading term of this series is considered. 
The core field is then equivalent to an elliptical line source
expansion and is fully characterized by the first moments 
of the line force distribution. 
Comparison with atomistic simulations have shown that this approach 
correctly describes the dislocation core field \cite{Gehlen1972,Hoagland1976,Woo1977,Henager2004,Henager2005}.

Until now, only few studies\cite{Gehlen1972,Kuan1976,Henager2004,Henager2005} 
have included this core field in the calculation of the elastic energy of dislocations
or of their  interaction with an external stress field.
Most of the time, dislocation elastic energies are obtained
by considering only the Volterra elastic field.
Such an approximated approach may lead to some errors.
Notably, simulation boxes used within \emph{ab initio} calculations
are usually too small to neglect the core field.
A recent study of the screw dislocation in iron \cite{Clouet2009b}
has shown, indeed, that it is necessary to include this core field
in the computation of the elastic energy when deriving 
core energies from \emph{ab initio} calculations.
The purpose of this paper is to extend the modeling
approach \cite{Hirth1973} of the core field within anisotropic 
linear elasticity so as to include its contribution in energy
calculations.
Previous studies, which have considered this core field contribution,
either assumed that the elastic behavior is isotropic
\cite{Gehlen1972,Kuan1976,Henager2004,Henager2005},
or that the elastic constants obey a given symmetry\cite{Hirth1973,Henager2005} 
incompatible for instance with a $\left<111\right>$ screw dislocation
in a cubic crystal.

In this paper, we first review how the elastic field of a line defect,
including the core field, is modeled within linear anisotropic elasticity theory. 
The approach of Hirth and Lothe \cite{Hirth1973} is generalized so as to describe
the core field not only through a distribution of line forces but also 
a distribution of dislocations.
The elastic energy of the line defect is then computed, 
thus showing the extra contribution arising from the core field.
Finally, we determine the influence of the core field 
on the interaction of the line defect with an external stress.

\section{Elastic field of a line defect}

We consider a static line defect, 
the line direction of which is denoted $\mathbf{e}_3$.
Such a defect can be either a dislocation, a line force, 
or the combination of both.
Eshelby \etal \cite{Eshelby1953} have shown that the elastic displacement 
and the stress created at a point of coordinates $\mathbf{x}$ 
can be, respectively, written as
\begin{subequations}
\label{eq:elas_esh}
\begin{align}
	u_k(\mathbf{x}) &= \frac{1}{2} \sum_{\alpha=1}^{6}{A_k^{\alpha} f_{\alpha}[z_{\alpha}]} , \\
	\sigma_{ij}(\mathbf{x}) &= \frac{1}{2} \sum_{\alpha=1}^{6}{
		B_{ijk}^{\alpha} A_k^{\alpha} \frac{\partial f_{\alpha}[z_{\alpha}]}{\partial z_{\alpha}}} ,
\end{align}
\end{subequations}
where the variable $z_{\alpha}$ is related to Cartesian coordinates 
by $z_{\alpha} = x_1 + p_{\alpha} x_2$.
The matrix $B_{ijk}^{\alpha}$ is obtained from elastic constants $C_{ijkl}$ by
\begin{equation*}
	B_{ijk}^{\alpha} = C_{ijk1} + p_{\alpha} C_{ijk2}.
\end{equation*}
The six roots $p_{\alpha}$ are the imaginary numbers, for which the following determinant 
is null
\begin{equation}
	\left|\left\{ B_{i1k}^{\alpha}+p_{\alpha}B_{i2k}^{\alpha} \right\}\right| = 0.
	\label{eq:sextic_p}
\end{equation}
The vectors $A_k^{\alpha}$, associated to each root $p_{\alpha}$,
are the non-null vectors that verify the relation
\begin{equation}
	(B_{i1k}^{\alpha}+p_{\alpha}B_{i2k}^{\alpha}) A_k^{\alpha} = 0.
	\label{eq:sextic_A}
\end{equation}
In all the above expressions and in the followings, we use the Einstein 
summation convention on repeated indexes, except for Greek indexes $\alpha$
that design the different roots $p_{\alpha}$. 
Such summations on $\alpha$ will always be explicitly indicated 
as in Eq.~(\ref{eq:elas_esh}).

The six roots $p_{\alpha}$ are necessary complex.
If $p_{\alpha}$ is a solution of Eq.~(\ref{eq:sextic_p}),
then its complex conjugate ${p_{\alpha}}^*$ is also a solution.
We sort the roots $p_{\alpha}$ according to the following 
usual rule
\begin{equation}
	\Im{(p_{\alpha})} > 0 \textrm{ and } p_{\alpha+3}={p_{\alpha}}^*,
	\quad \forall \alpha \in [1:3],
	\label{eq:p_cc}
\end{equation}
where $\Im{(p_{\alpha})}$ is the imaginary part of $p_{\alpha}$.
With such a convention, the matrices $B_{ijk}^{\alpha}$ verify the relation
\begin{equation}
	B_{ijk}^{\alpha+3} = {B_{ijk}^{\alpha}}^* ,
	\label{eq:Bijk_cc}
\end{equation}
and the vectors $A_k^{\alpha}$ can be chosen so that
\begin{equation}
	A_{k}^{\alpha+3} = {A_{k}^{\alpha}}^*.
	\label{eq:Ak_cc}
\end{equation}
As the elastic displacement has to be real, the functions $f_{\alpha}$
have also the property
\begin{equation}
	f_{\alpha+3}(z^*) = { f_{\alpha}(z) }^* .
	\label{eq:f_cc}
\end{equation}

The general form of the function $f_{\alpha}$ defining the elastic displacement 
and the stress (Eq.~\ref{eq:elas_esh}) is a Laurent series \cite{Eshelby1953}.
If we restrict ourselves to a line defect in an infinite crystal, 
the series is limited to the following terms
\begin{equation}
	f_{\alpha}(z) = \mp\frac{1}{2\pi\mi} \left( 
		D_{\alpha} \ln{(z)}
		+ \sum_{k=-\infty}^{1}{C^{k}_{\alpha}{z}^k} \right),
	\label{eq:Laurent_serie}
\end{equation}
with $\mi=\sqrt{-1}$.
The sign $\mp$ in this equation has to be taken as $-$ for $1\leq\alpha\leq3$
(roots having a positive imaginary part) and $+$ for $4\leq\alpha\leq6$
(roots having a negative imaginary part).
$\ln{(z)}$ is the principal determination of the complex logarithm,
whose imaginary part belongs to $[-\pi:\pi[$, thus showing a discontinuity 
in $\mathbb{R}^-$.

Far from the line defect, the main contribution to the function $f_{\alpha}$,
and thus to the elastic displacement, arises from the logarithm term. 
This corresponds to the Volterra elastic field created by a dislocation 
and to the 2D elastic Green function for a line force. 
Close to the line defect, the other terms in Eq. (\ref{eq:Laurent_serie})
may also lead to a relevant contribution. 
For a dislocation, these additional terms describe the core field. 
In the following, we only consider the main contribution to the core field
corresponding to the term $k=1$ in the Laurent series.
This correctly describes the core field far enough from the line defect.
The superposition of the Volterra and of the core fields, 
which gives the total elastic field created by a line defect, 
is then obtained from the following truncated series
\begin{equation*}
	f_{\alpha}(z) \underset{r\rightarrow\infty}{\sim} 
	\mp \frac{1}{2\pi\mi} \left( D_{\alpha} \ln{(z)}
		+ C_{\alpha}^{-1} \frac{1}{z} \right).
\end{equation*}

\subsection{Volterra elastic field}

The Volterra elastic field is given by the logarithm in 
Eq. (\ref{eq:Laurent_serie}). 
This leads to the following displacement and stress fields
\begin{subequations}
\label{eq:elast_esh_Volterra}
\begin{align}
	u_k^{\rm V}(\mathbf{x}) &= \frac{1}{2} \sum_{\alpha=1}^{6}{ 
		\mp \frac{1}{2\pi\mi} A_k^{\alpha} D_{\alpha}\ln{(z_{\alpha})}} , \\
		\sigma_{ij}^{\rm V}(\mathbf{x}) &= \frac{1}{2} \sum_{\alpha=1}^{6}{
		\mp \frac{1}{2\pi\mi} B_{ijk}^{\alpha} A_k^{\alpha} D_{\alpha}\frac{1}{z_{\alpha}}} .
\end{align}
\end{subequations}
This corresponds to the long range elastic field of a dislocation
of Burgers vector $\mathbf{b}$ or a line force of amplitude $\mathbf{F}$
if the coefficients $D_{\alpha}$ verify the system of equations \cite{Eshelby1953}
\begin{equation}
	\begin{split}
		\frac{1}{2} \sum_{\alpha=1}^{6}{A_k^{\alpha} D_{\alpha}} &= -b_k , \\
		\frac{1}{2} \sum_{\alpha=1}^{6}{ B_{i2k}^{\alpha} A_k^{\alpha} D_{\alpha}} &= -F_i.
	\end{split}
	\label{eq:D_eshelby}
\end{equation}

Stroh \cite{Stroh1958,Stroh1962} proposed a simple solution to this system of equations. 
In that purpose, he defined a new vector
\begin{equation}
	L_i^{\alpha} = B_{i2k}^{\alpha}A_k^{\alpha} 
	             = -\frac{1}{p_{\alpha}}B_{i1k}^{\alpha}A_k^{\alpha} .
	\label{eq:L_Stroh}
\end{equation}
As the vectors $A_i^{\alpha}$ are defined through the equation (\ref{eq:sextic_A}), 
their norm is not fixed. One can therefore choose their norm so that
\begin{equation}
	2 A_i^{\alpha} L_i^{\alpha} = 1, \quad \forall \alpha.
	\label{eq:normalisation_Stroh}
\end{equation}
Stroh showed that such a definition of the vectors $A_i^{\alpha}$
and $L_i^{\alpha}$ leads to the orthogonality property
\begin{equation*}
	A_i^{\alpha} L_i^{\beta} + A_i^{\beta} L_i^{\alpha} = \delta_{\alpha\beta},
\end{equation*}
where $\delta_{\alpha\beta}$ is the Kronecker symbol.
These vectors also verify the following relations \cite{Stroh1962,Bacon1980}
\begin{equation*}
	\sum_{\alpha=1}^6{A_i^{\alpha} A_j^{\alpha}} = 0, \quad
	\sum_{\alpha=1}^6{L_i^{\alpha} L_j^{\alpha}} = 0 \quad \textrm{and} \quad
	\sum_{\alpha=1}^6{A_i^{\alpha} L_j^{\alpha}} = \delta_{ij}.
\end{equation*}
These orthogonality properties lead to the expression of the coefficient $D_{\alpha}$:
\begin{equation}
	D_{\alpha} = -2 \left( L_i^{\alpha} b_i + A_i^{\alpha} F_i \right).
	\label{eq:D_Stroh}
\end{equation}

\subsection{Core field}

The Volterra solution models the elastic field created by a dislocation
far enough from the dislocation core. 
Close to the core, the dislocation core field may be relevant too.
We model this additional elastic field by considering the term $1/z$
in Eq. (\ref{eq:Laurent_serie}).
Gehlen \etal \cite{Gehlen1972} have shown that this term may be obtained
from dipoles of line forces. 
It is also possible to consider dipoles of dislocations,
which may be more natural to model the core field of a dissociated dislocation. 
Therefore, we assume that the core field can be modeled by 
an equilibrated distribution of dislocations and line forces 
of force amplitude $\mathbf{F}^q$ and of Burgers vector
$\mathbf{b}^q$ located at $\mathbf{a}^q$. 
All line force and dislocation directions are assumed to be collinear
to $\mathbf{e}_3$.
As the distribution is equilibrated, the resultant of the forces
and the total Burgers vector have to vanish
\begin{equation}
	\sum_q{\mathbf{F}^q}=\mathbf{0}
	\quad \textrm{and} \quad
	\sum_q{\mathbf{b}^q}=\mathbf{0}.
	\label{eq:equilibrium_distrib}
\end{equation}
The elastic displacement and the stress of this distribution
is given by the superposition of the Volterra elastic field created 
by each line defect
\begin{align*}
	u^{\rm c}_k(\mathbf{x}) &= \sum_q{u_k^{\textrm{V}^{(q)}}(\mathbf{x}-\mathbf{a}^q)} , \\
	\sigma^{\rm c}_{ij}(\mathbf{x}) &= \sum_q{\sigma_{ij}^{\textrm{V}^{(q)}}(\mathbf{x}-\mathbf{a}^q)} .
\end{align*}
We then assume that the norm of $\mathbf{x}$ is large compared to the norm
of the vectors $\mathbf{a}^q$.
One can thus make a limited expansion \cite{Gehlen1972,Hirth1973,Bacon1980,Ting1996}
leading to
\begin{align*}
	u^{\rm c}_k(\mathbf{x}) &= - \sum_q{ \frac{ \partial u_k^{\textrm{V}^{(q)}}(\mathbf{x}) }
		{\partial x_m} a^q_m } + O\left( \|\mathbf{a}^q\|^2 \right) , \\
	\sigma^{\rm c}_{ij}(\mathbf{x}) &= - \sum_q{ \frac{ \partial \sigma_{ij}^{\textrm{V}^{(q)}}(\mathbf{x}) } 
		{\partial x_m} a^q_m } + O\left( \|\mathbf{a}^q\|^2 \right),
\end{align*}
where we have used Eq.~(\ref{eq:equilibrium_distrib}) to eliminate 
the first term of the expansion.
Using Eq. (\ref{eq:elast_esh_Volterra}) 
and taking the limit $\mathbf{a}^q\to\mathbf{0}$, 
one finally obtains the expression of the core field
\begin{subequations}
	\label{eq:elasticity_couple}	
	\begin{align}
	u^{\rm c}_k(\mathbf{x}) &= \frac{1}{2} \sum_{\alpha=1}^{6}{ \mp \frac{1}{2\pi\mi}
		A_k^{\alpha} C^{-1}_{\alpha}
		\frac{1}{x_1 + p_{\alpha} x_2} }, \\
	\sigma^{\rm c}_{ij}(\mathbf{x}) &= \frac{1}{2} \sum_{\alpha=1}^{6}{ \pm \frac{1}{2\pi\mi}
		B_{ijk}^{\alpha} A_k^{\alpha} C^{-1}_{\alpha}
		\frac{1}{ \left( x_1 + p_{\alpha} x_2 \right)^2} } ,
	\end{align}
\end{subequations}
with
\begin{equation*}
	C^{-1}_{\alpha} = 2 A_i^{\alpha} \left( M_{i1} + p_{\alpha} M_{i2} \right)
	+ 2 L_i^{\alpha} \left( P_{i1} + p_{\alpha} P_{i2} \right),
\end{equation*}
where $M$ and $P$ are respectively the first moment tensors 
of the line force and of the dislocation distribution
\begin{equation*}
	M_{ij} = \sum_q{ F^q_i a^q_j }
	\quad \textrm{and} \quad
	P_{ij} = \sum_q{ b^q_i a^q_j }.
\end{equation*}
As we assume that the distribution of line defects representative
of the core field is equilibrated, it does not produce any torque.
This implies that the tensor $M_{ij}$ is symmetric\cite{Bacon1980}. 
The first moment tensors $M$ and $P$ can be simply deduced 
from the homogeneous stress computed in atomistic simulations using 
periodic boundary conditions\cite{Clouet2009b}.
Another method based on path-independent interaction integrals
computed through the field observed in atomistic simulations 
has also been proposed \cite{Soare2003,Soare2004}.

\section{Elastic energy of an isolated line defect}

We now calculate the elastic energy of a line defect, such as a dislocation,
taking into account its core field. The elastic field created by the line
defect is thus the superposition of the Volterra solution given by
Eq.~(\ref{eq:elast_esh_Volterra}) and of the core field given by 
Eq.~(\ref{eq:elasticity_couple}).
We define the elastic energy of the line defect as the integral 
of the elastic energy density over the volume 
in-between two cylinders centered on the line defect. 
The inner cylinder of radius $r_{\rm c}$ isolates the line defect core:
elastic fields are diverging at the line defect position and one needs 
to exclude the core region, where elasticity breaks down. 
The external cylinder of radius $R_{\infty}$ is introduced to prevent
the elastic energy from diverging.
Then, Gauss theorem allows us to obtain the elastic energy
\begin{equation}
	E = \frac{1}{2} \oint_S{\left( \sigma_{ij}^{\rm V} + \sigma_{ij}^{\rm c} \right)
	\left( u_i^{\rm V} + u_i^{\rm c} \right) \md S_j},
	\label{eq:elastic_energy}
\end{equation}
where the integration surface $S$ is composed of both cylinder surfaces 
and the branch cut, which isolates the displacement discontinuity.
We consider cylinders of unit height so as to express the elastic energy
per unit length of line defect.

This elastic energy can be decomposed into three different contributions:
the contribution of the Volterra solution, the contribution of the core field
and the cross interaction between both elastic fields. 

\subsection{Volterra contribution}

The Volterra contribution corresponds to the product 
$\sigma_{ij}^{\rm V}u_i^{\rm V}$ in Eq.~(\ref{eq:elastic_energy}).
It is given by the well-known result \cite{Foreman1955,Stroh1958,Stroh1962}
\begin{equation}
	E^{\rm V} = \frac{1}{2}\left( b_iK^0_{ij}b_j + F_iK'^0_{ij}F_j \right)
	\ln{\left( \frac{R_{\infty}}{r_{\rm c}} \right)},
	\label{eq:elastic_energy_Volterra}
\end{equation}
where we have defined the second rank tensors
\begin{equation}
	K^0_{ij}  = \sum_{\alpha=1}^6{ \pm\frac{1}{2\pi\mi} L_i^{\alpha}L_j^{\alpha} }
		\textrm{ and }
	K'^0_{ij}  = \sum_{\alpha=1}^6{ \mp\frac{1}{2\pi\mi} A_i^{\alpha}A_j^{\alpha} }.
	\label{eq:energy_prelog}
\end{equation}

\subsection{Core field contribution}
\label{sec:elastic_energy_core_field}

The contribution of the core field to the elastic energy corresponds 
to the product $\sigma_{ij}^{\rm c}u_i^{\rm c}$ in Eq.~(\ref{eq:elastic_energy}).
As the core field does not create any displacement discontinuity, 
the integration surface is simply composed of the inner and the external
cylinders. This leads to the contribution
\begin{multline*}
	E^{\rm c} = -\frac{1}{8}\sum_{\alpha=1}^{6}
	\mp\frac{1}{2\pi\mi}A_i^{\alpha}C_{\alpha}^{-1} \\
	\sum_{\beta=1}^{6} \pm\frac{1}{2\pi\mi}
	\left[B_{i1k}^{\beta}I_x^3(p_{\alpha},p_{\beta}) 
		+B_{i2k}^{\beta}I_y^3(p_{\alpha},p_{\beta}) \right] \\
	A_k^{\beta}C_{\beta}^{-1}
	\left( \frac{1}{{r_{\rm c}}^2} - \frac{1}{{R_{\infty}}^2} \right).
\end{multline*}
The integrals $I_x^3(p,q)$ and $I_y^3(p,q)$ are defined in the Appendix
(Eq.~(\ref{eq:integral_I3})).
We use the fact that $I_x^3(p,q)=-pI_y^3(p,q)$ 
as well as the property verified by the vectors $A_k^{\alpha}$ (Eq. \ref{eq:sextic_A}) 
and the definition of the vectors $L_i^{\alpha}$ (Eq. \ref{eq:L_Stroh}) 
to obtain
\begin{multline*}
	E^{\rm c} = -\frac{1}{32\pi^2} \sum_{\alpha=1}^6 \pm
	\sum_{\beta=1}^6 \pm C_{\alpha}^{-1} A_i^{\alpha} L_i^{\beta}C_{\beta}^{-1}
	\left[ p_{\alpha}p_{\beta}+1 \right] \\
	I_y^{3}(p_{\alpha},p_{\beta})
	\left( \frac{1}{{r_{\rm c}}^2} - \frac{1}{{R_{\infty}}^2} \right).
\end{multline*}
Finally, the expression of the integral $I^3_y(p,q)$ given in the appendix
allows us to write
\begin{multline}
	E^{\rm c} = \frac{1}{4\pi} \Im{\left( \sum_{\alpha=1}^3 \sum_{\beta=1}^3
	\frac{1+p_{\alpha}{p_{\beta}}^*}{(p_{\alpha}-{p_{\beta}}^*)^2}
	C_{\alpha}^{-1} A_i^{\alpha} {L_i^{\beta}}^* {C_{\beta}^{-1}}^* \right)} \\
	\left( \frac{1}{{r_{\rm c}}^2} - \frac{1}{{R_{\infty}}^2} \right).
	\label{eq:elastic_energy_core_field}
\end{multline}
The dependence of this expression with $R_{\infty}$ shows that the elastic
energy of the core field is concentrated close to the line defect. 
It is possible to take the limit $R_{\infty}\to\infty$, and thus to define
an elastic energy associated with the core field in the whole volume
excluding the core region, where elasticity breaks down.

\subsection{Volterra - core field interaction}
\label{sec:interaction_Volterra_core_field}

Then we calculate the interaction energy between both elastic fields
created by the line defect. Two different integrals can be used
to obtain this interaction energy \cite{Bacon1980}:
\begin{equation*}
	E^{\rm V-c} = \oint_S{\sigma_{ij}^{\rm V} u_i^{\rm c} \md S_j}
	            = \oint_S{\sigma_{ij}^{\rm c} u_i^{\rm V} \md S_j}.
\end{equation*}
We rather use the first definition to evaluate $E^{\rm V-c}$: 
as the core field displacement $\mathbf{u}^{\rm c}$ does not show
any discontinuity except at the origin, the integration surface 
of the first integral is simply composed of the inner and external 
cylinders. This leads to the following interaction energy
\begin{multline*}
	E^{\rm V-c} = \frac{1}{4}\sum_{\alpha=1}^6{\mp\frac{1}{2\pi\mi} A_i^{\alpha}C_{\alpha}^{-1}} \\
	\sum_{\beta=1}^6{\mp\frac{1}{2\pi\mi} \left( B_{i1k}^{\beta}I^2_x(p_{\alpha},p_{\beta})
	+ B_{i2k}^{\beta}I^2_y(p_{\alpha},p_{\beta}) \right) A_k^{\beta}D_{\beta}} \\
	\left( \frac{1}{r_{\rm c}} - \frac{1}{R_{\infty}} \right).
\end{multline*}
The integrals $I_{x}^2(p,q)$ and $I_{y}^2(p,q)$ are defined in the appendix
(Eq.~(\ref{eq:integral_I2})).
As these integrals vanish for any values of $p$ and $q$, this leads to 
$E^{\rm V-c} = 0$. 
As a result, there is no interaction energy between the 
Volterra elastic field and the core field of the line defect,
and the elastic energy of a line defect is simply the sum
of the elastic energies of the Volterra field and of the core field.
Of course, this is true only when the Volterra and the core fields
are centered at the same point. This may be imposed by symmetry,
as for the $\langle111\rangle$ screw dislocation in a cubic crystal.
\cite{Clouet2009b,Clouet2011b} 
When the Volterra and the core fields have different centers,
\cite{Gehlen1972,Hoagland1976,Henager2004,Henager2005} 
an interaction energy between both elastic fields exists.
Such a cross term can be simply calculated by considering 
the interaction of the core field with the stress created 
by the Volterra field, as described in the next section.

\section{Interaction with a stress field}

We now consider the interaction energy between an external stress field 
$\sigma_{ij}^{\rm ext}$ and a line defect.
The external stress can be an applied stress or the stress originating from another defect. 
The line defect is located at the origin and its line direction is $\mathbf{e}_3$.
It is characterized by its Burgers vector $\mathbf{b}$,
its force resultant $\mathbf{F}$ and the first moments tensors $M_{ij}$ and $P_{ij}$.
The interaction energy can be decomposed into two contributions:
the interaction with the Volterra elastic field and the interaction 
with the core field. 
The first contribution is well known.\cite{Hirth1982,Bacon1980}
For a dislocation, it is given 
by the integral of $\sigma_{ij}^{\rm ext}b_i\md{S_j}$ along the 
dislocation cut, where $\md{S_j}$ is an infinitesimal surface vector.
For a line force, it is given by the scalar product $F_iu_i^{\rm ext}$,
where $u_i^{\rm ext}$ is the displacement associated to the external stress field.
We now determine the contribution of the core field to the interaction energy $E_{\rm c}^{\rm inter}$.

\subsection{Core field contribution}

The interaction energy of the core field with the stress field $\sigma_{ij}^{\rm ext}$
can be obtained by considering the line defect distribution responsible for the core field,
thus using the same approach as used by Siems\cite{Siems1968,Bacon1980} for a point defect.
The interaction energy is then given by
\begin{equation*}
	E_{\rm c}^{\rm inter} = 
	\sum_q{ \int_0^1{ \sigma_{ij}^{\rm ext}(\lambda \mathbf{a}^q)
		b_i^q \epsilon_{jk3} a_k^q \md\lambda  }
	- F_i^q u_i^{\rm ext}(\mathbf{a}^q)},
\end{equation*}
where $\epsilon_{jkl}$ is the permutation tensor.
The first term represents the interaction with the different dislocations
of the distribution ($\epsilon_{jk3} a_k^q \md\lambda$ is the infinitesimal
surface vector along the dislocation cut),
and the second term represents the interaction with the line forces. 
A limited expansion of $\sigma_{ij}^{\rm ext}$ and of $u_i^{\rm ext}$
at the origin leads to
\begin{multline*}
	E_{\rm c}^{\rm inter} = 
	\sum_q \sigma_{ij}^{\rm ext}(\mathbf{0})
	b_i^q \epsilon_{jk3} a_k^q 
	- F_i^q \frac{\partial u_{i}^{\rm ext}(\mathbf{0})}{\partial x_j} a_j^q \\
	+ O\left( \|\mathbf{a}^q\|^2 \right) .
\end{multline*}
We finally use the fact that the tensor $M_{ij}$ is symmetric
and take the limit $\mathbf{a}^q \to \mathbf{0}$ to obtain 
the interaction energy
\begin{equation}
	E_{\rm c}^{\rm inter} = 
	\sigma_{ij}^{\rm ext}(\mathbf{0})\left( \epsilon_{jk3}P_{ik}
	- S_{ijkl} M_{kl}\right),
	\label{eq:Einter_core_field}
\end{equation}
where the elastic compliances $S_{ijkl}$ are the inverse of the elastic
constants ($S_{ijkl}C_{klmn}=\frac{1}{2}\left( \delta_{im}\delta_{jn} 
+ \delta_{in}\delta_{jm} \right)$).

Thus Eq.~(\ref{eq:Einter_core_field}) shows that an additional contribution
needs to be considered in the interaction energy of a line defect
with a stress when a core field is present.
In particular, this contribution of the core field leads to 
a dislocation-pressure interaction which can modify
the kink formation energy and the dislocation line tension 
at high pressures \cite{Hirth2005}.
Such a dependence of the dislocation energy with the pressure 
has been observed in atomistic simulations\cite{Pizzagalli2009b,Rabier2010,Yang2010}.
Eq.~(\ref{eq:Einter_core_field}) should allow to model this dependence,
or at least the first order variation.

\subsection{Force acting on a line defect}

The external stress field $\sigma_{ij}^{\rm ext}$ creates on the line defect
a force which derives from the interaction energy. 
Without the core field, this force would be simply given by
the Peach-Koehler formula for a pure dislocation 
and by the product $-S_{ijkl}\sigma_{kl}^{\rm ext}F_j$
for a pure line force.
Because of the core field, there is an additional force $\mathbf{f}^{\rm c}$
acting on the line defect. This force derives from the interaction
energy given by Eq.~(\ref{eq:Einter_core_field})
\begin{equation}
	f^{\rm c}_n = 
	- \frac{ \partial \sigma_{ij}^{\rm ext}(\mathbf{0}) }{\partial x_n}
	\left( \epsilon_{jk3}P_{ik}
	- S_{ijkl} M_{kl}\right).
	\label{eq:f_core_field}
\end{equation}
Because of the core field, there is a force acting on the line defect,
which is proportional to the gradient of the applied stress.

\subsection{Elastic energy of an isolated dipole}
\label{sec:elastic_energy_dipole}

The calculation of the elastic energy of an isolated dislocation dipole 
is one important application, where one needs to calculate the interaction 
energy of a line defect with a stress field. In that case,
the external stress field is created by the other line defect 
composing the dipole.
Here we determine the elastic energy of a dislocation dipole
that is assumed to be isolated from any other defect.
This elastic energy is defined as the integral of the energy density 
on the whole volume except two cylinders of radius $r_{\rm c}$ excluding
the regions around the dislocation core. As the elastic energy 
is now converging, we do not need to introduce an external cylinder
as we did for an isolated dislocation.
The dipole is composed of two line defects of opposite Burgers vector $\mathbf{b}$
and opposite force resultant $\mathbf{F}$ having the same core field
characterized by the moment tensors $M_{ij}$ and $P_{ij}$.
We assume that $\mathbf{e}_3$ corresponds to the line direction of the dislocations.
The dipole is then defined by the distance $d$ between the two dislocations
and by the angle $\phi$ between the dipole direction and a reference vector $\mathbf{e}_1$.

If the elastic field created by each dislocation composing the dipole
is only of the Volterra type [Eq.~(\ref{eq:elast_esh_Volterra})], 
the elastic energy of the dipole is
\begin{equation}
	E^{\rm V}_{\rm dipole} = \left( b_i K^0_{ij}b_j + F_i K'^0_{ij}F_j \right) 
		\ln{\left( \frac{d}{r_{\rm c}} \right)}
		+ 2 E^{\rm V}_{\rm c}(\phi),
	\label{eq:elastic_energy_dipole_Volterra}
\end{equation}
where the tensors $K^0_{ij}$ and $K'^0_{ij}$ are given by Eq.~(\ref{eq:energy_prelog}).
$E^{\rm V}_{\rm c}$ is the contribution from core tractions to the elastic energy. 
Such a contribution arises from the work done by the tractions of the Volterra elastic field
exerted on the cylinders that isolate the dislocation cores. 
It exists even when the core field is neglected and it is given by \cite{Clouet2009a}
\begin{widetext}
\begin{multline*}
	E_{\rm c}^{\rm V}(\phi) = \frac{1}{8}\sum_{\alpha=1}^6{
	\ln{(\mi \pm p_{\alpha})} 
	\sum_{\beta=1}^6{
	\pm\frac{1}{2\pi\mi}D_{\alpha}
	\left( A_i^{\alpha}L_i^{\beta}-L_i^{\alpha}A_i^{\beta} \right)
	D_{\beta}}} 
	+ \frac{1}{8\pi\mi}\sum_{\alpha=1}^3{\sum_{\beta=4}^6{
	D_{\alpha}
	\left( A_i^{\alpha}L_i^{\beta}-L_i^{\alpha}A_i^{\beta} \right)
	D_{\beta} \ln{\left( p_{\alpha}-p_{\beta} \right)} }} \\
	+ \frac{1}{2}\sum_{\alpha=1}^6{\pm\frac{1}{2\pi\mi}
	\left( b_iL_i^{\alpha}L_j^{\alpha}b_j
	- F_iA_i^{\alpha}A_j^{\alpha}F_j \right)
	\ln{(\cos{\phi}+p_{\alpha}\sin{\phi})}}.
	\label{eq:energy_dislo_core}
\end{multline*}
\end{widetext}

Considering now that a core field as described by Eq.~(\ref{eq:elasticity_couple})
is also created by each dislocation, one has to add to the elastic energy
of the dipole [Eq.~(\ref{eq:elastic_energy_dipole_Volterra})] 
the contribution from the core field of each dislocation, $2E^{\rm c}$,
as given by Eq.~(\ref{eq:elastic_energy_core_field}) in the limit
$R_{\infty}\to\infty$.

Another contribution also needs to be taken into account in the elastic energy
when dislocations create both a Volterra and a core field. 
It arises from the interaction of the total stress field created by the first
dislocation with the core field of the second one, and vice versa.
This interaction energy can be calculated using Eq.~(\ref{eq:Einter_core_field}),
which leads to the result
\begin{equation}
	E^{\rm V-c}_{\rm dipole} = 
	\left( 2\sigma^{\rm V}_{ij}(\mathbf{d}) + \sigma^{\rm c}_{ij}(\mathbf{d}) \right)
	\left( \epsilon_{jk3} P_{ik} - S_{ijkl} M_{kl} \right),
	\label{eq:elastic_energy_dipole_Volterra_core_field}
\end{equation}
where the vector $\mathbf{d}$
is defined by the coordinates $d(\cos{\phi},\sin{\phi},0)$.
Equation (\ref{eq:elastic_energy_dipole_Volterra_core_field}) shows 
that the elastic energy of the dipole now contains a contribution
varying with the inverse of the distance $d$ 
and another contribution varying with the square of the inverse of $d$.

\subsection{Dislocation dipole in periodic boundary conditions}

When studying dislocations in atomistic simulations, one can use
periodic boundary conditions. 
A dipole is introduced to ensure 
that the total Burgers vector of the simulation box is null.
Atomic simulations allow obtaining the excess energy 
associated with the defects present in the simulation box. 
One can deduce from this quantity dislocation intrinsic 
energy properties such as their core energy. To do so, one 
needs to calculate the elastic energy contained in the simulation
box. This elastic energy includes the elastic energy 
of the primary dipole as well as half the interaction energy
with all its periodic images.
When the simulation box is small, as in \emph{ab initio}
calculations, one needs to consider not only the Volterra field
but also the core field of the dislocations when computing 
the elastic energy\cite{Clouet2009b}.

The elastic energy of the primary dipole is given 
in the preceding section.
The interaction energy between two dipoles can be obtained
by decomposing it into the contributions arising from the different
constituents of the elastic field.
The interaction energy arising from the Volterra field of each dipole is obtained 
thanks to the expression given by Stroh \cite{Stroh1958} for the interaction
energy between two dislocations. 
If the coordinates of the vectors joining the two dislocations 
are $(x_1,x_2,x_3)$, this part of the interaction energy is given by
\begin{equation*}
	E_{\rm inter}^{\rm V-V} = 
	- \sum_{\alpha=1}^6{ \pm \frac{1}{2\pi\mi} 
	b_i^{(1)}L_i^{\alpha}L_j^{\alpha}b_j^{(2)}
	\ln{\left( x_1 + p_{\alpha}x_2 \right)}},
\end{equation*}
where $\mathbf{b}^{(1)}$ and $\mathbf{b}^{(2)}$ are 
the respective Burgers vectors of each dislocation\footnote{We 
assume pure dislocations and therefore consider that there is 
no line force: $\mathbf{F}^{(1)}=\mathbf{F}^{(2)}=0$.}.

The part of the interaction energy arising from the core field
is obtained thanks to Eq.~(\ref{eq:Einter_core_field}). 
The external stress $\sigma_{ij}^{\rm ext}$ appearing in this equation 
corresponds to the stress created by the other dislocations,
where both the Volterra and the core fields are considered.

When summing all contributions from the different periodic images, 
one should be aware that the sums are only conditionally convergent.
This convergence problem can be easily resolved using the regularization
method of Cai \etal \cite{Cai2003}.

\section{Conclusions}

We have extended in this paper the approach of Hirth and Lothe \cite{Hirth1973}
to model dislocation core fields within linear anisotropic elasticity theory
by deriving the elastic energy of a straight dislocation  
while taking full account of its core field.
The obtained expression shows that this elastic energy is the sum 
of the energies corresponding to the Volterra field and to the core field, 
and that no cross interaction exists between these two elastic fields.
We have also shown that the core field leads to an additional contribution
to the interaction energy between a dislocation and an external stress.
Through this contribution, the energy of a dislocation 
can depend on the applied pressure.
This interaction with the core field is also responsible
for an additional force acting on the dislocation, which is proportional
to the gradient of the applied stress.
Dislocation properties may therefore be affected in regions where 
a high-stress gradient is present such as in a dislocation pile-up
\cite{Kuan1976,Hirth1993b} or close to a crack tip.

The interaction of the dislocations caused by their core field
is shorter-range than their interaction through their Volterra field.
It will therefore affect the interaction between dislocations
when they get close enough. 
Such a situation may arise in atomistic simulations, where the size 
of the simulation box may be too small to neglect the influence 
of the core field.\cite{Clouet2009b}
One should then take account of the dislocation core field
when calculating elastic energies, 
which can be done using the different
expressions of this paper.
An example is given in the following paper,\cite{Clouet2011b}
where the developed formalism is applied to the 
$\langle111\rangle$ screw dislocation in $\alpha$-iron.

\begin{acknowledgments}
	The author thanks Lisa Ventelon for her careful reading of the manuscript.
\end{acknowledgments}

\appendix*

\section{Integrals}

The elastic energy of the core field (\cf §\ref{sec:elastic_energy_core_field})
makes the two following integrals appear
\begin{equation}
	\begin{split}
	I_{x}^3(p,q) &= \int_{-\pi}^{\pi}{\frac{\cos{\theta}}
		{(\cos{\theta}+p\sin{\theta})(\cos{\theta}+q\sin{\theta})^2}\md\theta}, \\
	I_{y}^3(p,q) &= \int_{-\pi}^{\pi}{\frac{\sin{\theta}}
	{(\cos{\theta}+p\sin{\theta})(\cos{\theta}+q\sin{\theta})^2}\md\theta}.
	\end{split}
	\label{eq:integral_I3}
\end{equation}
These integrals of a rational function of $\cos{(\theta)}$ and $\sin{(\theta)}$
can be evaluated using the residues theorem \cite{Arfken2001}. 
This leads to the result
\begin{align*}
	I_{x}^3(p,q) 	
		&= 0	\phantom{\frac{4\pi\mi p}{(p-q)^2}}			
			& \textrm{ if } \Im{(p)}>0 \textrm{ and } \Im{(q)}>0,\\
		&= \frac{4\pi\mi p}{(p-q)^2}	\phantom{0}
			& \textrm{ if } \Im{(p)}>0 \textrm{ and } \Im{(q)}<0,\\
		&= \frac{-4\pi\mi p}{(p-q)^2}	\phantom{0}
			& \textrm{ if } \Im{(p)}<0 \textrm{ and } \Im{(q)}>0,\\
		&= 0 	\phantom{\frac{4\pi\mi p}{(p-q)^2}}
			& \textrm{ if } \Im{(p)}<0 \textrm{ and } \Im{(q)}<0,\\
\end{align*}
\begin{align*}
	I_{y}^3(p,q) 
		&= 0  	\phantom{\frac{4\pi\mi p}{(p-q)^2}}
			& \textrm{ if } \Im{(p)}>0 \textrm{ and } \Im{(q)}>0, \\
		&= \frac{-4\pi\mi}{(p-q)^2} \phantom{0}
			& \textrm{ if } \Im{(p)}>0 \textrm{ and } \Im{(q)}<0, \\
		&= \frac{4\pi\mi}{(p-q)^2} 	\phantom{0}
			& \textrm{ if } \Im{(p)}<0 \textrm{ and } \Im{(q)}>0, \\
		&= 0  	\phantom{\frac{4\pi\mi p}{(p-q)^2}}
			& \textrm{ if } \Im{(p)}<0 \textrm{ and } \Im{(q)}<0. \\
\end{align*}

The two integrals appearing in the interaction energy between the Volterra
and the core fields of a line defect (\cf §\ref{sec:interaction_Volterra_core_field})
are
\begin{equation}
	\begin{split}
	I_{x}^2(p,q)  &= \int_{-\pi}^{\pi}{\frac{\cos{\theta}}
		{(\cos{\theta}+p\sin{\theta})(\cos{\theta}+q\sin{\theta})}\md\theta} \\
	I_{y}^2(p,q)  &= \int_{-\pi}^{\pi}{\frac{\sin{\theta}}
		{(\cos{\theta}+p\sin{\theta})(\cos{\theta}+q\sin{\theta})}\md\theta}
	\end{split}
	\label{eq:integral_I2}
\end{equation}
The residues theorem leads to the result $I_{x}^2(p,q) = I_{y}^2(p,q) = 0$.

\bibliographystyle{apsrev4-1}
\bibliography{clouet2011a}

\end{document}